\documentclass{article}

\usepackage{arxiv}

\usepackage[utf8]{inputenc} % allow utf-8 input
\usepackage[T1]{fontenc}    % use 8-bit T1 fonts
\usepackage{hyperref}       % hyperlinks
\usepackage{url}            % simple URL typesetting
\usepackage{booktabs}       % professional-quality tables
\usepackage{amsfonts}       % blackboard math symbols
\usepackage{nicefrac}       % compact symbols for 1/2, etc.
\usepackage{microtype}      % microtypography
\usepackage{lipsum}
\usepackage{graphicx}
\graphicspath{ {./images/} }

\usepackage{cite}
\usepackage{amsmath,amssymb,amsfonts}
\usepackage{textcomp}
\usepackage{comment}
\usepackage{algorithm}
\usepackage{algorithmic}

\usepackage{ragged2e}

\title{Characterizing the Interactions Between Classical and Community-aware Centrality Measures in Complex Networks}

\author{
  Stephany Rajeh*, Marinette Savonnet, Eric Leclercq, and Hocine Cherifi \\
  Laboratoire d'Informatique de Bourgogne\\
  University of Burgundy\\
  21000 Dijon, France\\
  \texttt{*stephany.rajeh@u-bourgogne.fr} \\
}

\begin{document}
\maketitle

\begin{abstract}
Identifying vital nodes in networks exhibiting a community structure is a fundamental issue. Indeed, community structure is one of the main properties of real-world networks. Recent works have shown that community-aware centrality measures compare favorably with classical measures agnostic about this ubiquitous property. Nonetheless, there is no clear consensus about how they relate and in which situation it is better to use a classical or a community-aware centrality measure. To this end, in this paper, we perform an extensive investigation to get a better understanding of the relationship between classical and community-aware centrality measures reported in the literature. Experiments use artificial networks with controlled community structure properties and a large sample of real-world networks originating from various domains. Results indicate that the stronger the community structure, the more appropriate the community-aware centrality measures. Furthermore, variations of the degree and community size distribution parameters do not affect the results. Finally, network transitivity and community structure strength are the most significant drivers controlling the interactions between classical and community-aware centrality measures.

\end{abstract}

% keywords can be removed
%\keywords{Hierarchy \and Centrality \and Complex Networks \and  Influential nodes}

\section*{Introduction}
Interactions between entities are pervasive in social, technological, infrastructural, information, and biological systems. Identifying influential nodes in those networks is a crucial problem. Indeed, multitude of applications exist compassing from combating epidemic outbreaks \cite{wang2017vaccination}, detecting essential proteins \cite{jalili2016evolution}, predicting contagions in animal groups \cite{rosenthal2015revealing}, to estimating robustness of infrastructure networks \cite{wandelt2018robustness}, planning landscapes \cite{de2016urban}, improving routing efficiency on the internet \cite{wang2019betweenness}, understanding information diffusion \cite{guille2013information} and many other more \cite{das2018study, lu2016vital}.  Centrality measures are one of the main approaches to deal with this issue. Classically, they quantify the importance of a node based on a typical network topological property \cite{das2018study, lu2016vital}. They can be classified into local and global measures depending on the topological information they process. Local measures rely on a node's ability to influence its neighborhood, while global measures are concerned with the ability of a node to influence the whole network. Generally, local measures require a low computation cost, while global ones are computationally intensive. More recent works tend to consider centrality as a multidimensional issue where both local and global information can be combined \cite{sciarra2018change, ibnoulouafi2018m}. 

While there is a great deal of work on designing centrality measures, the mainstream does not exploit the network's community structure. However, it is a ubiquitous property observed in the vast majority of real-world networks \cite{fortunato2016community,jebabli2014overlapping,jebabli2015overlapping}. A community is generally apprehended as a group of nodes densely connected between each other and sparsely connected with the other nodes of the network \cite{newman2004finding}. As communities play a significant role in understanding how nodes behave in networks \cite{riolo2020consistency, girvan2002community, mantzaris2014uncovering}, a research area concerned with the relation between community structure and the importance of nodes has recently emerged in network science. These works have shown that incorporating community structure information allows designing more effective centrality measures \cite{ghalmane2019immunization, ghalmane2019centrality, hwang2006bridging, tulu2018identifying, gupta2016centrality, zhao2015community, guimera2005functional, luo2016identifying}. We refer to them as ``community-aware'' centrality measures. One can divide them into three groups (local, global, and mixed measures) according to the type of link they consider. Indeed, in modular networks, one can distinguish intra-community links from inter-community links. Intra-community links (strong ties/short-range interactions) connect nodes belonging to the same community, while inter-community links (weak ties/long-range interactions) connect nodes belonging to different communities \cite{granovetter1977strength}. Local measures exploit intra-community links. In other words, local measures view the neighborhood as the dimension to identify influential nodes. Global measures target inter-community links. In this case, influential nodes are the bridges connecting different communities. Finally, mixed measures target influential nodes based on a combination of their local and global characteristics.

This study investigates ten classical (community-agnostic) centrality measures and twenty-eight community-aware centrality measures. This representative set covers the spectrum of the various categories, and it contains the most influential ones.
The local classical centrality measures are Degree, Leverage, Laplacian, Diffusion Degree, and Maximum Neighborhood Component. The global classical centrality measures are Betweenness, Closeness, Katz, PageRank, and Subgraph centrality.
Among the twenty-eight community measures, twenty are based on the "modular centrality" proposed by Ghalmane \textit{et al.}\cite{ghalmane2019centrality}. In this work, the authors claim that a node centrality has a local and a global component in modular networks. The local dimension is computed on the network extracted from the original modular network by removing the inter-community links. The global dimension is computed on the network obtained by keeping the inter-community links and the nodes attached to these links in the original network. Consequently, one can extend any classical centrality measure to its so-called modular version.  For example, the local degree centrality of a node is the number of its intra-community links. Its global component is the number of its inter-community links. The ten local community-aware centrality measures under investigation are the local component of the modular centrality derived from the ten classical centrality measures \cite{ghalmane2019centrality}. The twelve global community-aware centrality measures are the Number of Neighboring Communities centrality \cite{ghalmane2019immunization}, the Bridging centrality \cite{hwang2006bridging} and the global component of the modular centrality of the ten classical centrality measures \cite{ghalmane2019centrality}.
Finally, the set of six mixed community-aware centrality is made of  Comm centrality\cite{gupta2016centrality}, Community‑based Mediator centrality\cite{tulu2018identifying}, Community Hub-Bridge centrality\cite{ghalmane2019immunization}, Community-based centrality\cite{zhao2015community}, Participation coefficient\cite{guimera2005functional}, and K-shell with Community centrality\cite{luo2016identifying}. 

%Our goal is to investigate the interplay between classical and community-aware centrality measures and the influence of the network topological properties on their interactions. 

Previous works investigated the relations between community structure and macroscopic network topological properties. In \cite{orman2011accuracy}, Orman \textit{et al.} show that the community structure strength significantly affects the average distance, transitivity, and assortativity of the network. Wharrie \textit{et al.}\cite{wharrie2019micro} conclude that high clustering naturally yields to a greater number of communities. Orman \textit{et al.} \cite{orman2013empirical} show that community structure strength and transitivity are positively correlated. Furthermore, a modular structure can still exist even with a low value of transitivity. The work by Lancichinetti \textit{et al.} \cite{lancichinetti2010characterizing} show that a set of mesoscopic characteristics such as community size distribution and average path length of networks originating from the same domain are relatively similar. A paper by Wang \textit{et al.} \cite{wang2010impact} reveals that as the community structure strength increases, the communicability in a network decreases. According to the work of Nematzadeh, \textit{et al.}  \cite{nematzadeh2014optimal}, the emergence of a global diffusion occurs after reaching a minimum threshold of community structure strength. All these works allow a better understanding of the relations between the community structure and the network's macroscopic topological properties. However, they do not give a clue on their effect on centrality measures.

Another set of works is concerned with the effect of network topology on classical centrality measures. Li \textit{et al.}\cite{li2015correlation} show that classical centrality measures are generally positively correlated and that correlation is independent of the network size. Ronqui \& Travieso\cite{ronqui2015analyzing} demonstrate that correlation values of classical centrality measures are higher on scale-free models as compared to in real-world networks. Schoch \textit{et al.}  \cite{schoch2017correlations} show that classical centrality measures are well correlated on threshold graphs. In \cite{rajeh2020interplay}, Rajeh \textit{et al.} report that density and transitivity significantly affect the correlation between centrality and hierarchy measures. Finally, Oldham \textit{et al.} \cite{oldham2019consistency} show that modularity is the main parameter driving the correlation between classical centrality measures. Although these works light up the relationship between network topology and classical centrality measures, they bring no information about community-aware centrality measures. To our knowledge, up to now, the relationship between classical and community-aware centrality measures and their relation with the network's topological property is still unexplored. An extensive study is performed on the interplay between classical and community-aware centrality measures to fill this gap. The influence of macroscopic and mesoscopic topological properties of the network are examined. This work allows answering questions such as in which practical situation community-aware centrality measures present an added-value as compared to classical centrality measures.

The main findings of this paper are as follows: \newline
1) As the communities become more and more well separated, the correlation values observed between global community-aware centrality measures and classical centrality measures decrease. Conversely, the correlation values between local community-aware centrality measures and classical ones increase. Mixed community-aware centrality measures are split into two categories. Some behave similarly to local community-aware centrality measures while the others behave like global community-aware centrality measures depending on which part they favor.\newline
2) Results are generally insensitive to the variations of the parameters of the degree and community size distribution.\newline
3)  Transitivity and the mixing parameter play a crucial role in driving the correlation variation between classical and community-aware centrality measures.\newline

\section*{Comparative evaluation using a synthetic network benchmark}

The correlation between classical and community-aware centrality measures is investigated using synthetic networks generated with the LFR model\cite{lancichinetti2008benchmark}. It allows controlling the strength of the community structure through the mixing parameter ($\mu$). This parameter is the ratio of the inter-community links to the total number of links in a network. The power-law exponents of the degree distribution ($\gamma$) and the community size distribution ($\theta$) are also tunable. A set of networks has been generated to evaluate these three parameters' influence on the Kendall's Tau correlation between all possible combinations of classical and community-aware centrality measures. The synthetic network parameters and their respective values are reported in \autoref{LFRtable}.

\subsection*{Influence of the community structure strength}
The mixing parameter ($0\leq\mu\leq 1$) controls the community structure strength. A low value of $\mu$ characterizes networks with a strong or well-defined community structure (very few inter-community links). By contrast, a high mixing parameter value results in a network with a weak community structure (a higher proportion of inter-community links than intra-community links). Nine networks spanning from very strong ($\mu$ = 0.05) to very weak ($\mu$ = 0.70) community structure are used in the experiments. Other parameters are  fixed to typical values ($\gamma$ = 2.7, $\theta$ = 2.7).

\begin{figure}[t]
\begin{center}
\includegraphics[width=1\textwidth,  height=8.6cm, keepaspectratio]{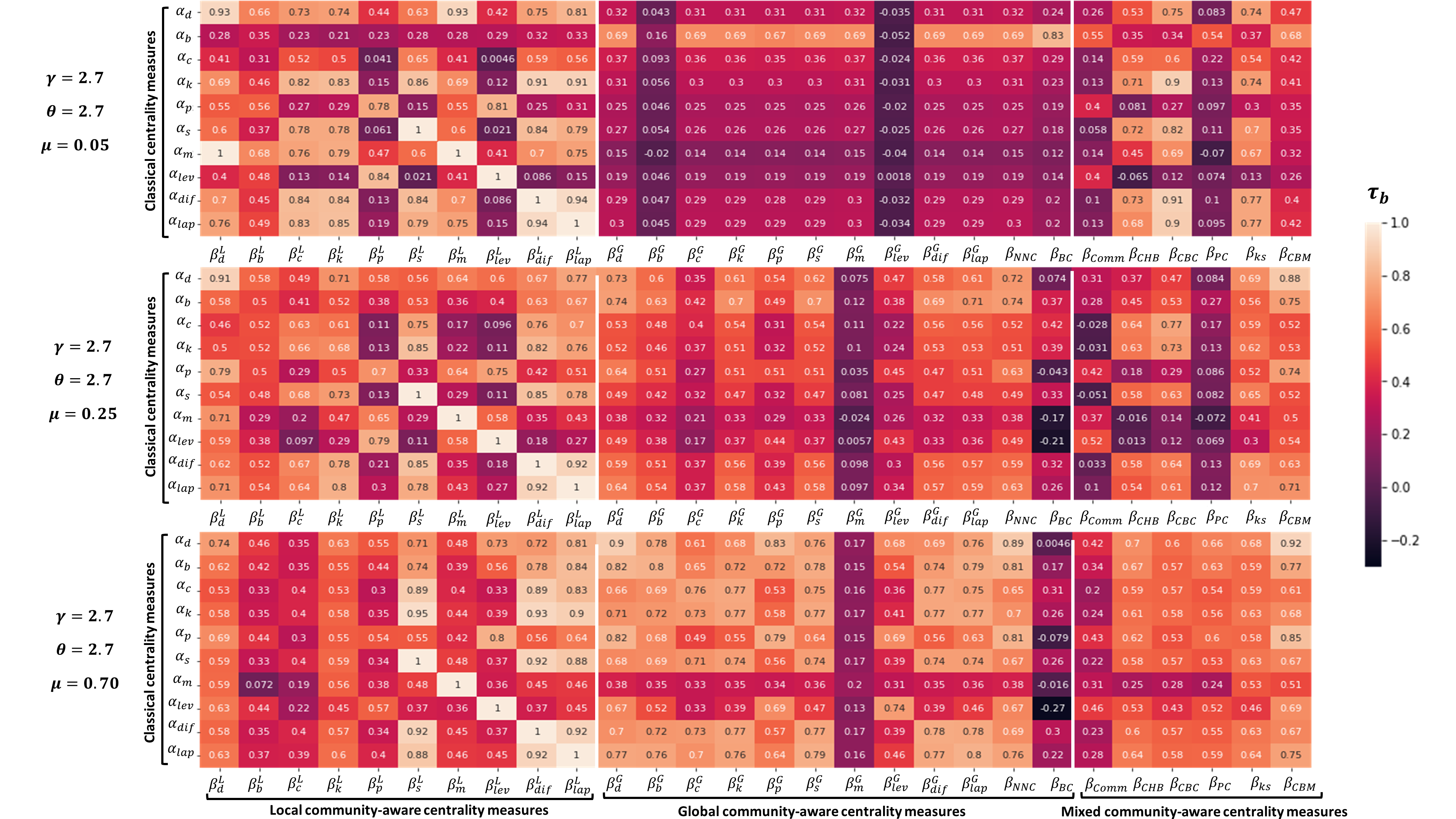}
 \caption{Heatmaps of Kendall’s Tau correlation of the various combinations between classical ($\alpha_i$) and community-aware ($\beta_j$) centrality measures in synthetic networks. $\gamma$ is the exponent of the degree distribution. $\theta$ is the exponent of the community size distribution. Three values of the mixing parameter  $\mu$ are reported [0.05, 0.25, 0.70]. The classical centrality measures are: $\alpha_d$ = Degree, $\alpha_b$ = Betweenness, $\alpha_c$ = Closeness, $\alpha_k$ = Katz, $\alpha_p$ = PageRank, $\alpha_s$ = Subgraph, $\alpha_m$ = Maximum Neighborhood Component, $\alpha_{lev}$ = Leverage, $\alpha_{dif}$ = Diffusion, $\alpha_{lap}$ =  Laplacian. The local community-aware centrality measures are: ($\beta^L_d$, $\beta^L_b$, $\beta^L_c$, $\beta^L_k$ ,$\beta^L_p$, $\beta^L_s$, $\beta^L_m$, $\beta^L_{lev}$, $\beta^L_{dif}$, $\beta^L_{lap}$) = the local component of the classical centrality measures based on modular centrality. The global community-aware centrality measures are: ($\beta^G_d$, $\beta^G_b$, $\beta^G_c$, $\beta^G_k$ ,$\beta^G_p$, $\beta^G_s$, $\beta^G_m$, $\beta^G_{lev}$, $\beta^G_{dif}$, $\beta^G_{lap}$) = the global component of the classical centrality measures based on modular centrality, $\beta_{NNC}$ = Number of Neighboring Communities centrality, $\beta_{BC}$ = Bridging centrality. The mixed community-aware centrality measures are: $\beta_{Comm}$  = Comm centrality, $\beta_{CHB}$  = Community Hub-Bridge centrality, $\beta_{CBC}$  = Community-based centrality, $\beta_{PC}$  = Participation Coefficient, $\beta_{ks}$ = K-shell with Community centrality,  $\beta_{CBM}$ = Community-based Mediator centrality.} 
 \label{SyntheticMainPaperBasicConfiguration}
\end{center}
\end{figure}

\begin{figure}[t]
\begin{center}
\includegraphics[width=1\textwidth,  height=8cm, keepaspectratio]{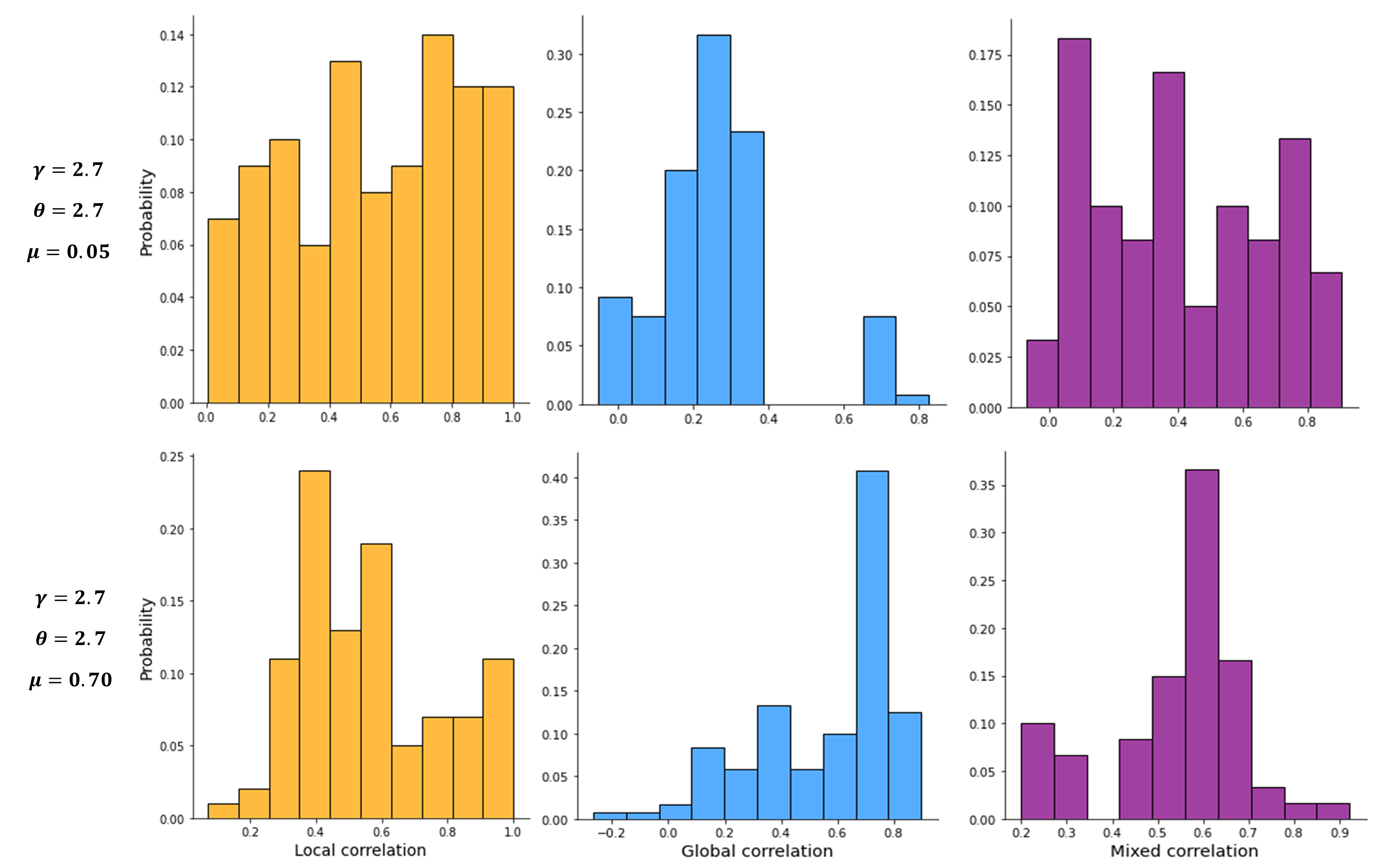}
 \caption{Histograms of the correlation between classical and local, global, and mixed community-aware centrality measures. The top  histograms are for a strong community structure ($\mu$ = 0.05). The bottom  histograms are for a weak community structure ($\mu$ = 0.70).}
  \label{HistogramsSynthetic}
\end{center}
\end{figure}

\begin{figure}[t]
\begin{center}
\includegraphics[width=1\textwidth,  height=5cm, keepaspectratio]{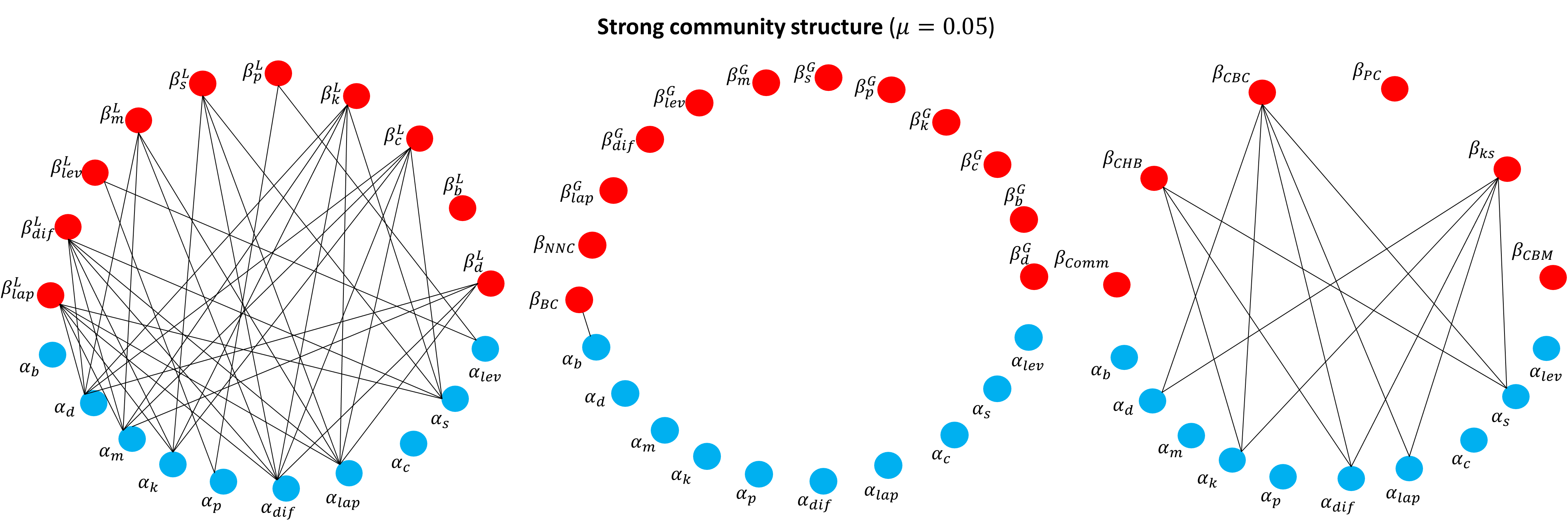}
\includegraphics[width=1\textwidth,  height=5cm, keepaspectratio]{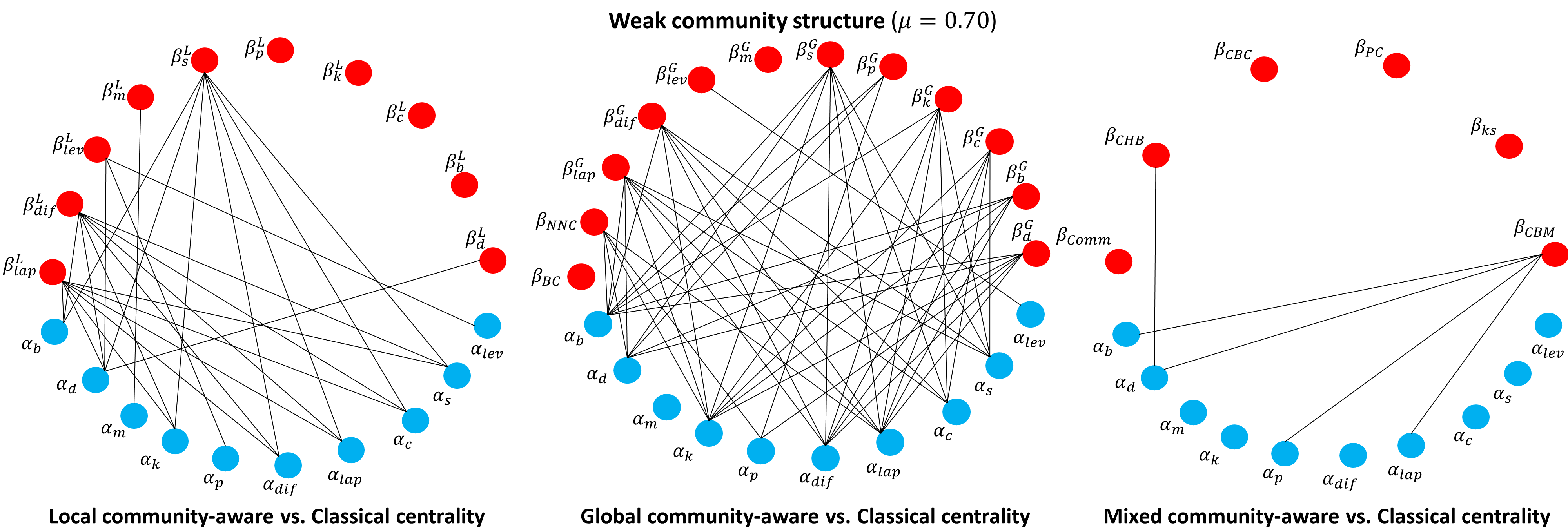}
 \caption{Network visualization of the correlation between classical ($\alpha_i$) and community-aware ($\beta_j$) centrality measures. Two nodes are connected if their Kendall's Tau correlation value is greater than 0.70. The blue nodes represent classical centrality measures. The red nodes represent community-aware centrality measures. The top networks are based on an LFR network generated with a strong community structure ($\mu$ = 0.05). The bottom networks are based on an LFR network generated with a weak community structure ($\mu$ = 0.70).}
  \label{NetworkVisualizationOfCorrelation}
\end{center}
\end{figure}

The heatmaps in \autoref{SyntheticMainPaperBasicConfiguration} represent the Kendall's Tau correlation between classical and community-aware centrality measures. The heatmap at the top is for a network with a strong community structure ($\mu = 0.05$). The middle concerns a network with a medium community structure strength ($\mu = 0.25$). Finally, the bottom is for a network with a weak community structure ($\mu = 0.70$). Fig. S1, in the supplementary material, contains the figures with other values of the community structure strengths ($\mu= 0.1, 0.15, 0.20, 0.30, 0.35, 0.40$). The heatmaps are arranged into three blocks. They correspond to local community-aware centrality measures, global community-aware centrality, and mixed community-aware centrality measures from left to right.

\begin{table}[t]
   \centering
    \caption{Synthetic network parameters generated by the LFR model.}
    \label{LFRtable}    
      \begin{tabular}{lc}
      \hline
      Network Parameter & Value \\
      \hline 
    %\hline
    Number of nodes & 2500 \\
    Average degree & 8 \\
    Maximum degree & 27 \\	
    Exponent for degree distribution ($\gamma$) & [2, 2.7, 3] \\	
    Exponent for community size distribution ($\theta$) & [2, 2.7, 3] \\
    Minimum community size &  4 \\
    Maximum community size & 250 \\
    Mixing parameter ($\mu$) & [0.05, 0.10, 0.15, 0.20, 0.25, 0.30, 0.35 0.40, 0.70] \\
    \hline
      \end{tabular}
\end{table}

% Strong --> Discuss global, then local, then mixed
For a strong community structure ($\mu$ = 0.05), the left block of the heatmaps representing the correlation between local community-aware centrality measures and the classical centrality measures is very patchy. It is a sign of wide variations. For example, the correlation value of the local component of PageRank modular centrality ($\beta_p^L$) with Closeness centrality ($\alpha_c$) is 0.041, while its correlation with Leverage centrality ($\alpha_{lev}$) reaches 0.84. The middle block of the heatmaps concerning the correlation between global community-aware centrality measures and classical centrality measures is more uniform. The vast majority of the observed correlation values are low. Finally, two groups can be extracted from the heatmaps on the right concerning the mixed community-aware centrality measures. The first behaves as the global community-aware centrality measures. Comm centrality ($\beta_{Comm}$), Participation Coefficient ($\beta_{PC}$) Community‑based Mediator centrality ($\beta_{CBM}$) fall into this group. The second behaves as the local community-aware centrality measures. Community Hub-Bridge centrality ($\beta_{CHB}$), Community-based centrality ($\beta_{CBC}$), and K-shell with Community centrality ($\beta_{ks}$) fall into this group.

% Explaining why
The reason why community-aware centrality measures can be divided into two groups is to be looked for in their definitions. Comm centrality ($\beta_{Comm}$) assumes that a node may act as a hub and as a bridge simultaneously. Additionally, bridges are given more weight under the assumption that they are generally rare. In a strong community structure, there are rare connections between nodes from different communities. Accordingly, if a node has external connections (i.e., acting as a bridge), it scores high in Comm centrality. This information is combined with the node's ability to be a hub in its community. As a result, Comm centrality exhibits low correlation with classical centrality measures that don't differentiate between inter-community and intra-community links. Participation coefficient ($\beta_{PC}$) assumes that the more a node has links linked externally as compared to its total links, the more important it is. This is why, in a network with strong community structure, the rare external links of nodes are highly accounted for. As a result, they show low correlation with classical centrality measures. Finally, Community‑based Mediator centrality ($\beta_{CBM}$) is based on the simultaneous entropy of the intra-community and inter-community links of a node. The subsequent entropy is then weighted by the total ratio of connections a node has in a network. Consequently, if a node acquires more inter-community links than intra-community links, it will be considered more important. On the other hand, if a node has an equal proportion of intra-community and inter-community links, $\beta_{CBM}$ of the node will turn into its normalized degree centrality. %Reptition: Participation coefficient ($\beta_{PC}$) assumes that the more a node has external links, the more critical it is. Therefore, in a network with strong community structure, the rare external links are highly valued. As a result, it exhibits a low correlation with classical centrality measures. Finally, Community‑based Mediator centrality ($\beta_{CBM}$) is based on the simultaneous entropy of the intra-community and inter-community links of a node. The subsequent entropy is then weighted by the total ratio of connections of the node. Consequently, if a node acquires more inter-community links than intra-community links, it is considered more important. 
The second group, which shows a high correlation with some classical centrality measures, is unusual. Community Hub-Bridge centrality ($\beta_{CHB}$) weights the node's intra-community links by the size of its community and inter-community links by the number of neighboring communities (i.e., the total number of communities a node has access to). If the number of communities a node can reach is similar across nodes, or if community sizes are comparable, $\beta_{CHB}$ reduces to degree centrality. In strong community structured networks, the number of communities a node can reach is not uniform across nodes since many communities are barely connected. Nonetheless, communities may be of comparable sizes. This may be the reason why $\beta_{CHB}$ behavior fluctuates. For example, it has a high correlation (0.71) with Katz centrality ($\alpha_k$). At the same time, it has a low correlation value (-0.065) with Leverage centrality ($\alpha_{lev}$). Community-based centrality ($\beta_{CBC}$) also weights the node's inter-community links with the community sizes. However, this time, the community size of the node's neighbors is taken into consideration, not the node's community size. The community size of the node is also considered to weight the intra-community links. Hence, this centrality is also sensitive to the community sizes. If the communities are of comparable sizes, $\beta_{CBC}$ behaves as Degree centrality. Finally, K-shell with Community centrality ($\beta_{ks}$) behaves similarly to Community Hub-Bridge centrality ($\beta_{CHB}$) and to Community-based centrality ($\beta_{CBC}$) although its definition differs. First, the original network is divided into a local network and a global network. Then, the K-shell hierarchical decomposition is performed. A node is then assigned to values, weighted by a parameter $\alpha$ to give preference to hubs in the local component as compared to bridges in the global component. It is set to 0.5 in this study (i.e., equal preference for hubs and bridges). It seems that a community hierarchical decomposition approach to the network doesn't always provide unique information from the whole network. It is interesting to note that only for the second group, Betweenness, PageRank, and Leverage centrality, are the only classical centrality measures showing a low correlation.

% Medium --> Discuss global, then local, then mixed

For a medium community structure strength ($\mu$ = 0.25), the values of the correlation between the local community-aware centrality measures and the classical centrality measures still fluctuate in a wide range. The global community-aware centrality measures show a slightly higher correlation than in the case of a strong community structure ($\mu$ = 0.05). For the mixed community-aware centrality measures, one can still form two groups behaving like the local and the global community-aware centrality measures. However, Community‑based Mediator centrality ($\beta_{CBM}$) shows now medium to high correlation with classical centrality measures. It is therefore very sensitive to a small change in the community structure strength.

% Weak --> Discuss global, then local, then mixed
For a weak community structure ($\mu$ = 0.70), the local community-aware centrality measures are less correlated to classical centrality measures. On the contrary, the global community-aware centrality measures exhibit higher correlation values.  Finally, the mixed community-aware centrality measures show medium to high correlation.  Comm centrality ($\beta_{Comm}$) still has a low correlation with classical centrality measures.
%?????Note that Maximum Neighborhood Component centrality calculated on the global component ($\beta^G_{m}$) and Bridging centrality ($\beta_{BC}$) still show low correlation even on a network exhibiting a weak community structure. To a less extent is Leverage centrality calculated on the global component ($\beta^G_{lev}$).????????

% Comment on the histograms 
\autoref{HistogramsSynthetic} shows the histograms of the correlation values for the three groups of community-aware centrality measures (local, global, mixed) considering the two extreme cases of the networks community structure strength (weak and strong). 
We concentrate on the modal-class quantifying the most frequent correlation range to compare these histograms according to the community structure strength. Let’s first look at the local community-aware centrality measures. The modal-class interval is [0.70; 0.80] in the strong community structure case ($\mu$ = 0.05). It shifts to [0.35; 0.45] in the weak community structure case ($\mu$ = 0.70). This behavior clearly illustrates that correlation decreases as the community structure get weaker for local community-aware centrality measures. This phenomenon is reversed for global community-aware centrality measures. Indeed, for a strong community structure ($\mu$ = 0.05), the modal-class interval is [0.2; 0.3] while it is [0.7; 0.8] for a weak community structure ($\mu$ = 0.70). For mixed community-aware centrality measures with strong community structure ($\mu$ = 0.05), the modal-class interval is [0.01; 0.10]. However, the dispersion of the correlation values is high. With a weak community structure ($\mu$ = 0.70), the modal-class interval becomes [0.55; 0.65] and correlation values become more concentrated at the modal-class interval. In other words, community-aware centrality measures exhibit more similar behavior when the community structure gets weaker.

 \autoref{NetworkVisualizationOfCorrelation} illustrates the correlation between classical and community-aware centrality measures based on a bipartite network representation. The classical centrality measures are the blue nodes, and the red nodes represent the community-aware centrality measures. Two nodes are connected if their correlation exceeds a threshold value of 0.70. For networks with a strong community structure ($\mu$ = 0.05), almost all local community-aware centrality measures correlate well with the classical centrality measures. In networks with a weak community structure ($\mu$ = 0.70), the network is sparser, signaling less frequent high correlation values. One can observe an opposite behavior for the global community-aware centrality measures. Indeed, there is a single link between Bridging centrality ($\beta_{BC}$) and Betweenness centrality ($\alpha_b$) in the strong community structure ($\mu$ = 0.05) case, while the network is dense in the weak community structure situation ($\mu$ = 0.70). High correlation values are less frequent between mixed community-aware and classical centrality measures. For networks with a strong community structure ($\mu$ = 0.05), K-shell with Community centrality ($\beta_{ks}$) and Community-based centrality ($\beta_{CBC}$) show high correlation with only four classical centrality measures. Namely, Degree centrality ($\alpha_d$), Katz centrality ($\alpha_k$), Diffusion centrality ($\alpha_{diff}$) and Laplacian centrality ($\alpha_{lap}$). Community Hub-Bridge centrality ($\beta_{CHB}$) shows a high correlation with the same classical centrality measures except for Degree centrality ($\alpha_d$). 
For networks with a weak community structure ($\mu$ = 0.70), correlation with classical centrality is high for only two community-aware centrality measures. Community Hub-Bridge centrality ($\beta_{CHB}$) shows a high correlation with four classical centrality measures. Namely, Betweenness centrality ($\alpha_b$), Degree centrality ($\alpha_d$), Katz centrality ($\alpha_p$), and Laplacian centrality ($\alpha_{lap}$). Besides, Community-based Mediator centrality ($\beta_{CBM}$) shows a high correlation only with degree centrality ($\alpha_d$). Therefore, mixed community-aware centrality measures generally do not highly correlate with classical centrality measures in a weak community structure.

Let’s summarize the main results of this experiment. First of all, the correlation values between the local community-aware centrality measures and the classical centrality measures range from medium to high in a network with a strong community structure. In conjunction, the global community-aware centrality measures exhibit a low correlation with classical centrality measures. Conversely, when a network has a weak community structure, the behavior of the global and local community-aware centrality measures is inverted. Mixed community-aware centrality measures can be divided into two groups when a network exhibits a strong community structure. One behaves as the local community-aware centrality measures and the other one as the global ones. In the case of a weak community structure, however, they generally demonstrate a medium correlation.

\subsection*{Influence of the degree distribution}

\begin{figure}[t!]
\begin{center}
\includegraphics[width=1\textwidth,  height=8.6cm, keepaspectratio]{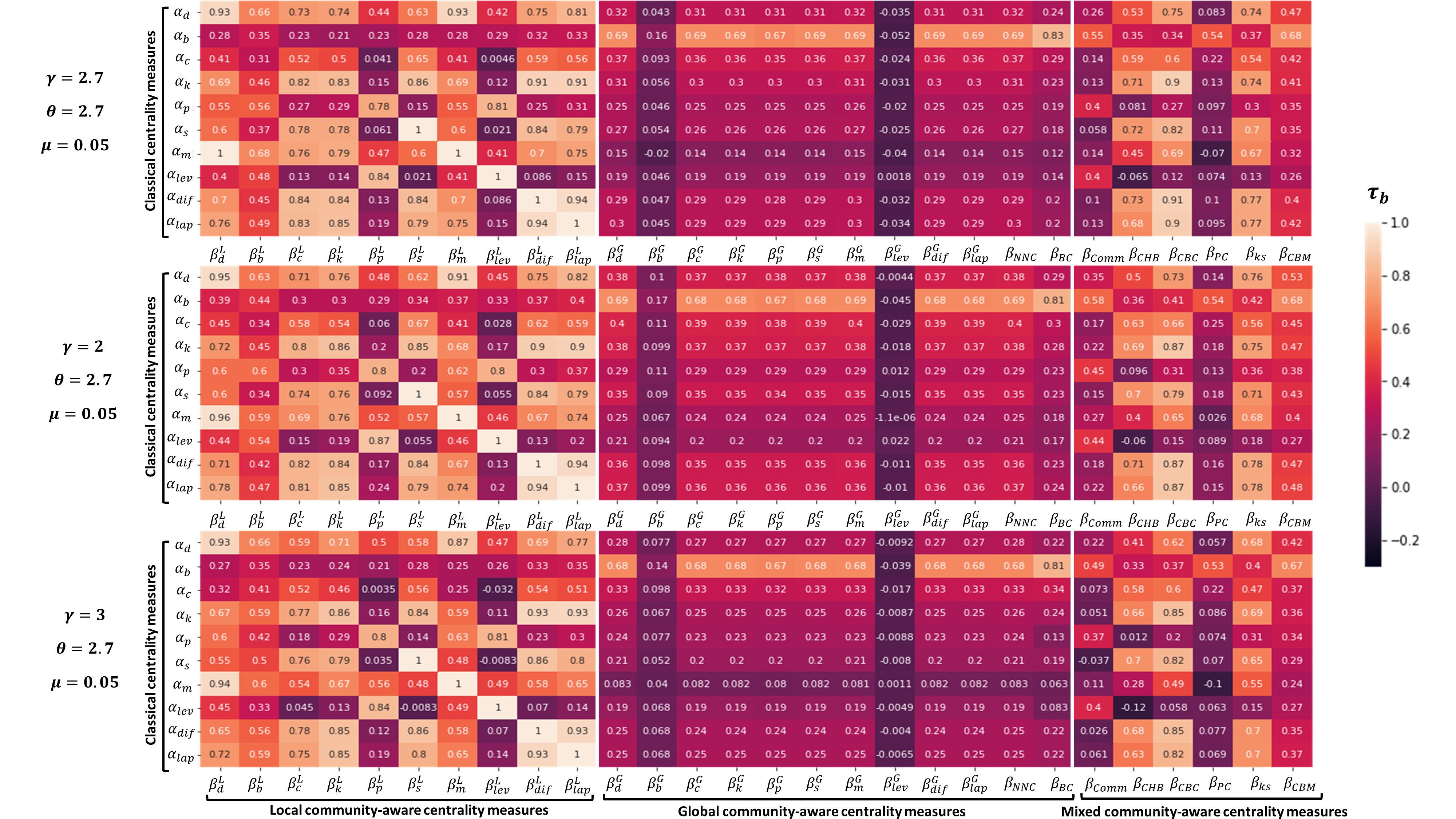}
 \caption{Heatmaps of Kendall’s Tau correlation of the various combinations between classical ($\alpha_i$) and community-aware ($\beta_j$) centrality measures in synthetic networks. $\gamma$ is the exponent of the degree distribution. Three values are used [2, 2.7, 3]. $\theta$ is the exponent of the community size distribution, and $\mu$ is the mixing parameter. The classical centrality measures are: $\alpha_d$ = Degree, $\alpha_b$ = Betweenness, $\alpha_c$ = Closeness, $\alpha_k$ = Katz, $\alpha_p$ = PageRank, $\alpha_s$ = Subgraph, $\alpha_m$ = Maximum Neighborhood Component, $\alpha_{lev}$ = Leverage, $\alpha_{dif}$ = Diffusion, $\alpha_{lap}$ =  Laplacian. The local community-aware centrality measures are: ($\beta^L_d$, $\beta^L_b$, $\beta^L_c$, $\beta^L_k$ ,$\beta^L_p$, $\beta^L_s$, $\beta^L_m$, $\beta^L_{lev}$, $\beta^L_{dif}$, $\beta^L_{lap}$) = the local component of the classical centrality measures based on modular centrality. The global community-aware centrality measures are: ($\beta^G_d$, $\beta^G_b$, $\beta^G_c$, $\beta^G_k$ ,$\beta^G_p$, $\beta^G_s$, $\beta^G_m$, $\beta^G_{lev}$, $\beta^G_{dif}$, $\beta^G_{lap}$) = the global component of the classical centrality measures based on modular centrality, $\beta_{NNC}$ = Number of Neighboring Communities centrality, $\beta_{BC}$ = Bridging centrality. The mixed community-aware centrality measures are: $\beta_{Comm}$  = Comm centrality, $\beta_{CHB}$  = Community Hub-Bridge centrality, $\beta_{CBC}$  = Community-based centrality, $\beta_{PC}$  = Participation Coefficient, $\beta_{ks}$ = K-shell with Community centrality,  $\beta_{CBM}$ = Community-based Mediator centrality.} 
 \label{SyntheticDegreeExpVariation}
\end{center}
\end{figure}

The goal of this experiment is to study the influence of the degree distribution. Indeed, the LFR algorithm allows tuning the power-law degree distribution exponent ($\gamma$) of the generated networks. Three values of the exponent ($\gamma$) are used.  The typical value ($\gamma$ = 2.7) from the previous experiment because it covers a large spectrum of real-world situations \cite{lancichinetti2008benchmark}.  A low value ($\gamma$ = 2.0) resulting in networks close to hub-and-spoke networks. Indeed, in this case, nodes' tendency to link to highly connected nodes is enhanced, accelerating the rich-gets-richer process \cite{barabasi2016network}. A high value ($\gamma$ = 3) where the rich-gets-richer phenomenon is less pronounced, such that the network characteristics are closer to a random network\cite{barabasi2016network, newman2003structure}. Like in the previous experiment, the Kendall's Tau correlation between the classical and the community-aware centrality measures is computed using networks with community structure strength ranging from strong ($\mu$ = 0.05) to weak ($\mu$ = 0.70). \autoref{SyntheticDegreeExpVariation} shows the heatmaps related to the three values of the degree distribution exponent using networks with a strong community structure ($\mu$ = 0.05). At first sight, the results are very similar. 

% Case 1: Hub-and-spoke + strong community structure

The Pearson correlation between the hub-and-spoke type network ($\gamma$ = 2) and the reference network ($\gamma$ = 2.7) heatmaps are computed considering the local, global and mixed community-aware centrality measures separately. Results confirm that the relation between classical and community-aware centrality measures is quite comparable. Indeed, correlation ranges from 0.986 to 0.994.
% Case 2: Random-like + strong community structure
For a random-like structure network ($\gamma$ = 3), the behavior is still relatively similar compared to the reference case ($\gamma$ = 2.7). The Pearson correlation between the two networks for local community-aware centrality measures amounts to 0.976. For global community-aware centrality measures, it is 0.984. Finally, for mixed community-aware centrality measures, it amounts to 0.987.

% Case 3: Hub-and-spoke + weak community structure
The heatmaps using networks with a weak community structure ($\mu$ = 0.70) are provided in Supplementary material Fig. 2. Results are in the same vein as previously. Indeed, varying the degree exponent ($\gamma$ = [2, 2.7, 3]) does not affect the heatmaps. For networks with a hub-and-spoke like structure ($\gamma$ = 2.7), the Pearson correlation between the various community-aware heatmaps and their counterparts in the reference case is also relatively high. It is equal to 0.971 for the local community-aware centrality measures, 0.992 for the global community-aware centrality measures and 0.893 for the mixed community-aware centrality measures. % Case 4: Random-like + weak community structure
The overall behavior is also similar for networks with a random-like structure ($\gamma$ = 3). Indeed, for local community-aware centrality measures, the Pearson correlation value is 0.984, and for the global community-aware centrality measures, it amounts to 0.990. Nevertheless, some differences are worth noticing in the mixed community-aware centrality measures set. Indeed, the Pearson correlation value is much lower (0.843). Community Hub-Bridge centrality ($\beta_{CHB})$, Community-based centrality ($\beta_{CBC}$), Participation coefficient ($\beta_{PC})$, and K-shell with Community centrality ($\beta_{ks}$) show a lower correlation with the classical centrality measures as compared to the reference network.

To summarize, varying the degree distribution exponent doesn't significantly affect the relationship between classical and community-aware centrality measures on artificial networks. However, there is a slight decrease in correlation for mixed community-aware centrality measures when the network tends to a random-like structure ($\gamma$ = 3). It is valid for strong and weak community structured networks.

\subsection*{Influence of the community size distribution}
In this experiment, we investigate the effect of community size distribution variations. The community size distribution of the networks generated by LFR is a power-law. Its exponent ($\theta$) is tunable. Like in the previous experiments, three values are considered.  With a small value ($\theta$ = 2), networks have many small communities coexisting with very few big communities. The reference case ($\theta$ = 2.7) also has many small communities coexisting with very few big communities. However, the proportion of small communities is less. Finally, networks have communities with comparable sizes for a high exponent value ($\theta$ = 3) \cite{barabasi2016network}. 

\begin{figure}[t!]
\begin{center}
\includegraphics[width=1\textwidth,  height=8.6cm, keepaspectratio]{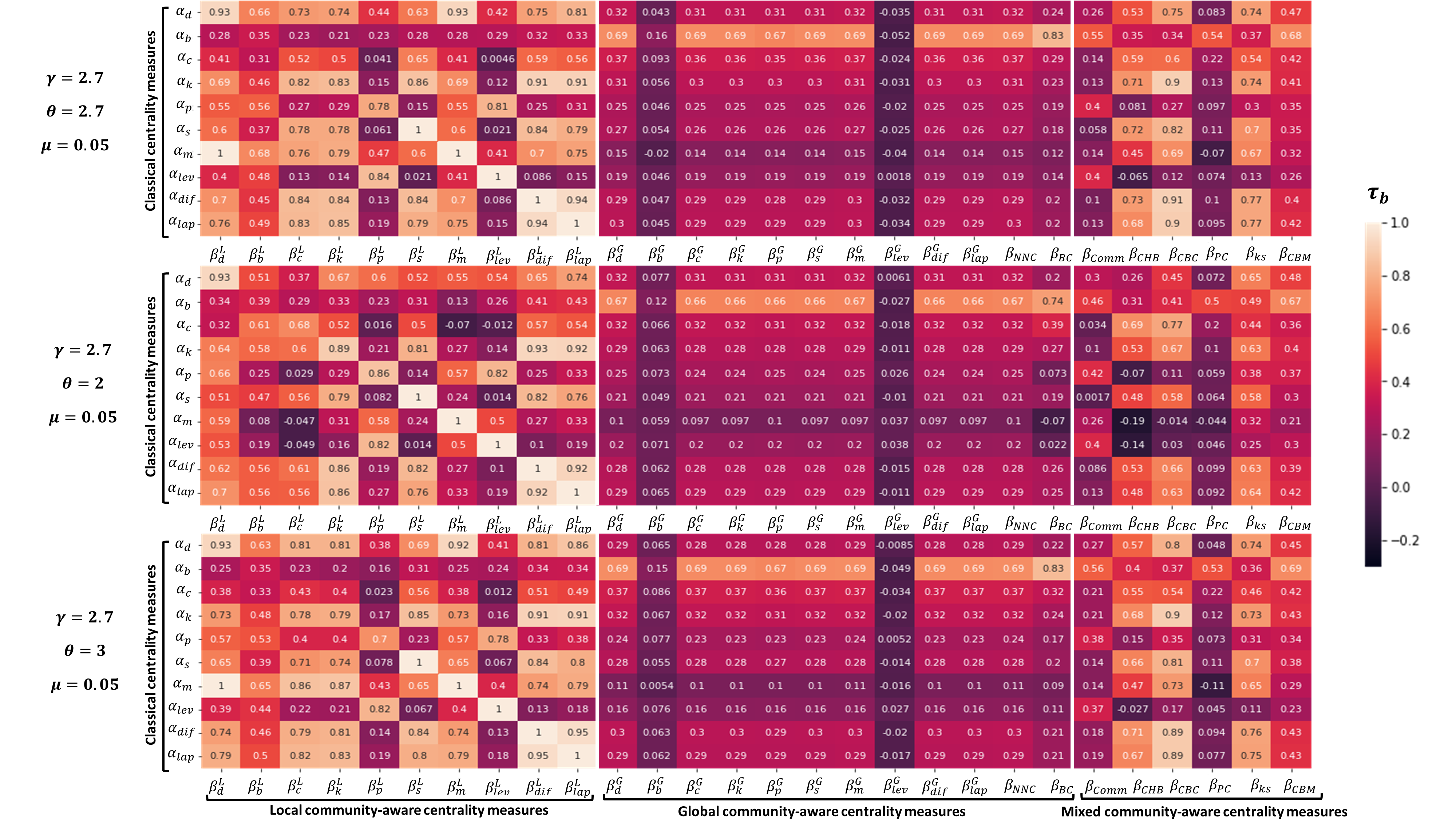}
 \caption{Heatmaps of Kendall’s Tau correlation of the various combinations between classical ($\alpha_i$) and community-aware ($\beta_j$) centrality measures in synthetic networks. $\gamma$ is the exponent of the degree distribution. $\theta$ is the exponent of the community size distribution. Three values are used [2, 2.7, 3]. $\mu$ is the mixing parameter. The classical centrality measures are: $\alpha_d$ = Degree, $\alpha_b$ = Betweenness, $\alpha_c$ = Closeness, $\alpha_k$ = Katz, $\alpha_p$ = PageRank, $\alpha_s$ = Subgraph, $\alpha_m$ = Maximum Neighborhood Component, $\alpha_{lev}$ = Leverage, $\alpha_{dif}$ = Diffusion, $\alpha_{lap}$ =  Laplacian. The local community-aware centrality measures are: ($\beta^L_d$, $\beta^L_b$, $\beta^L_c$, $\beta^L_k$ ,$\beta^L_p$, $\beta^L_s$, $\beta^L_m$, $\beta^L_{lev}$, $\beta^L_{dif}$, $\beta^L_{lap}$) = the local component of the classical centrality measures based on modular centrality. The global community-aware centrality measures are: ($\beta^G_d$, $\beta^G_b$, $\beta^G_c$, $\beta^G_k$ ,$\beta^G_p$, $\beta^G_s$, $\beta^G_m$, $\beta^G_{lev}$, $\beta^G_{dif}$, $\beta^G_{lap}$) = the global component of the classical centrality measures based on modular centrality, $\beta_{NNC}$ = Number of Neighboring Communities centrality, $\beta_{BC}$ = Bridging centrality. The mixed community-aware centrality measures are: $\beta_{Comm}$  = Comm centrality, $\beta_{CHB}$  = Community Hub-Bridge centrality, $\beta_{CBC}$  = Community-based centrality, $\beta_{PC}$  = Participation Coefficient, $\beta_{ks}$ = K-shell with Community centrality,  $\beta_{CBM}$ = Community-based Mediator centrality.} 
 \label{SyntheticCommunitySizeExpVariation}
\end{center}
\end{figure}

% Case 1: High proportion of small communities + strong community structure

\autoref{SyntheticCommunitySizeExpVariation} shows heatmaps of networks having a strong community structure ($\mu$ = 0.05)  for the three values of the community size distribution exponent. At first glance, they all seem alike. For networks with a high proportion of small communities coexisting with few large communities ($\theta$ = 2), the Pearson correlation between the local heatmap and its counterpart in the reference ($\theta$ = 2.7) is the lowest (0.790). Indeed, for a group of local community-aware centrality measures, correlation with the classical centrality measures decreases. For global community-aware centrality measures, the Pearson correlation value is the highest (0.976), reflecting the similarity with the reference. Finally, with a value of 0.819, the Pearson correlation is for the mixed community-aware centrality measures. % Case 2:  Communities of comparable sizes + strong community structure
For networks with comparable community sizes ($\theta$ = 3), the correlation between classical and community-aware centrality measures is very similar to the reference case ($\theta$ = 2.7). Indeed, the Pearson correlation is high in the three categories of community-aware centrality measures. It amounts to 0.987 for the local community-aware centrality measures, to 0.993 for the global community-aware centrality measures, and 0.990 for the mixed community-aware centrality measures. 

% Case 3:  High proportion of small communities + weak community structure
Let’s now examine the case of networks generated with a weak community structure strength ($\mu$ = 0.70) with varying the community size distribution exponent. Heatmaps are available in Supplementary Fig. S3.
For networks with a high proportion of small communities coexisting with few large communities ($\theta$ = 2), they generally show similar behavior that the reference case ($\theta$ = 2.7). Indeed, the Pearson correlation between the network at hand ($\theta$ = 2.7) and the reference case ($\theta$ = 2) is high, with a value of 0.956 for local community-aware centrality measures and 0.992 for global community-aware centrality. Nevertheless, one can observe differences with mixed community-aware centrality measures. The Pearson correlation value is equal to 0.733 in this case. For example, Comm centrality ($\beta_{Comm}$) shows a higher correlation than the reference case for all classical centrality measures. It could happen because Comm centrality ($\beta_{Comm}$) relies on bridges and hubs, while small communities may lack both and have comparable degrees. As a result, it shows a higher correlation with classical centrality measures. % Case 4:  Communities of comparable sizes + weak community structure
For networks with communities of comparable sizes ($\theta$ = 3), the heatmaps are very similar to the reference case ($\theta$ = 2.7). In every case, the Pearson correlation between the heatmaps is relatively high. It amounts to 0.993 for the local community-aware centrality measures, to 0.996 for the global community-aware centrality measures, and 0.949 for the mixed community-aware centrality measures. 

To summarize, varying the community size distribution exponent does not significantly affect the relationship between classical and community-aware centrality measures. Yet, in networks with a high proportion of small communities ($\theta$ = 2), whatever the community structure strength, mixed community-aware centrality measures tend to have a lower correlation with classical centrality than in the reference case ($\theta$ = 2.7). 

\section*{Comparative evaluation using real-world networks}

Extensive experiments are performed on a set of 50 real-world networks originating from various fields (social, biological, infrastructural, collaboration, etc.)  \cite{lusseau2003bottlenose, nr, icon, latora2017complex, rozemberczki2020characteristic, netz, kunegis2014handbook}. First of all, as there is no ground truth, the community structure is uncovered. The correlation between classical and community-aware centrality measures is computed using Kendall's Tau, and the influence of the community structure strength is investigated. The community detection is performed using Infomap \cite{rosvall2008maps} as a reference and Louvain \cite{blondel2008fast}. The goal is to investigate the sensitivity of the results to the community structure variation associated with the community detection algorithm. Finally, the correlation values are further processed to relate them with the macroscopic and mesoscopic properties of the networks using linear regression. Supplementary Note 4 provides a brief description of the real-world networks. 

\subsection*{Influence of the community structure strength}
\begin{figure}[t!]
\begin{center}
\includegraphics[width=1\textwidth, height=8.6cm, keepaspectratio]{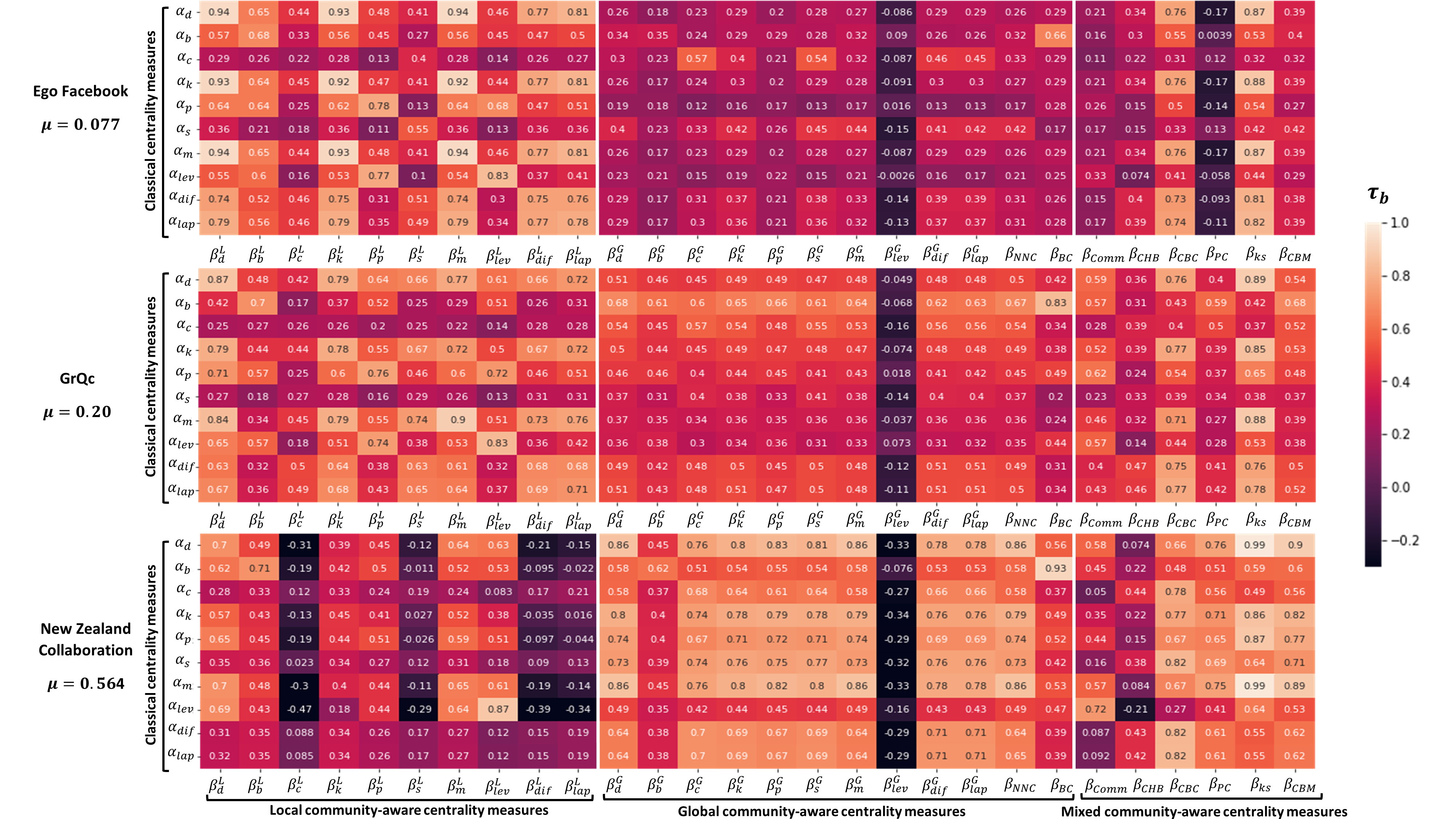}
 \caption{Heatmaps of Kendall’s Tau correlation of the various combinations between classical ($\alpha_i$) and community-aware ($\beta_j$) centrality measures in 3 real-world networks. Networks are sorted in ascending order according to their mixing parameter ($\mu$). The mixing parameter is deduced after the community structure is uncovered using Infomap community detection algorithm. The classical centrality measures are: $\alpha_d$ = Degree, $\alpha_b$ = Betweenness, $\alpha_c$ = Closeness, $\alpha_k$ = Katz, $\alpha_p$ = PageRank, $\alpha_s$ = Subgraph, $\alpha_m$ = Maximum Neighborhood Component, $\alpha_{lev}$ = Leverage, $\alpha_{dif}$ = Diffusion, $\alpha_{lap}$ =  Laplacian. The local community-aware centrality measures are: ($\beta^L_d$, $\beta^L_b$, $\beta^L_c$, $\beta^L_k$ ,$\beta^L_p$, $\beta^L_s$, $\beta^L_m$, $\beta^L_{lev}$, $\beta^L_{dif}$, $\beta^L_{lap}$) = the local component of the classical centrality measures based on modular centrality. The global community-aware centrality measures are: ($\beta^G_d$, $\beta^G_b$, $\beta^G_c$, $\beta^G_k$ ,$\beta^G_p$, $\beta^G_s$, $\beta^G_m$, $\beta^G_{lev}$, $\beta^G_{dif}$, $\beta^G_{lap}$) = the global component of the classical centrality measures based on modular centrality, $\beta_{NNC}$ = Number of Neighboring Communities centrality, $\beta_{BC}$ = Bridging centrality. The mixed community-aware centrality measures are: $\beta_{Comm}$  = Comm centrality, $\beta_{CHB}$  = Community Hub-Bridge centrality, $\beta_{CBC}$  = Community-based centrality, $\beta_{PC}$  = Participation Coefficient, $\beta_{ks}$ = K-shell with Community centrality,  $\beta_{CBM}$ = Community-based Mediator centrality.} 
  \label{RealNetworksInfomapMainPaper}
\end{center}
\end{figure}

In this experiment, the correlation between classical and community-aware centrality measures is compared according to the community structure strength.  As the community structure of the real-world networks is unknown, Infomap is used to uncover it. It allows estimating their mixing parameter value ($\mu$) and classifying them into three groups. The first group contains networks with a strong community structure. They have a mixing parameter value in the range $\mu$ = [0.05; 0.19]. A mixing parameter value $\mu$ = [0.20; 0.30] characterizes networks with a medium community structure strength. Finally, for networks with a weak community structure strength, the mixing parameter value fall in the range of $\mu$ = [0.30, 1].  We illustrate our findings with a typical example of each class in \autoref{RealNetworksInfomapMainPaper}. The supplementary material contains the results for the other networks. Of course, we report about networks that deviate from the typical behavior of their group.

Ego Facebook is a typical network with a strong community structure ($\mu$ = 0.077). The heatmap of local community-aware centrality measures shows high variability. For example, the correlation between Degree centrality based on intra-community links ($\beta_d^L$) and Katz centrality ($\alpha_k$) is high (0.93), while it is low (0.29) with Closeness centrality ($\alpha_c$). Results are more homogeneous for global community-aware centrality measures. The correlation with classical centrality measures is generally low. One can observe these two types of behavior for the mixed community-aware centrality measures. Indeed, Comm centrality ($\beta_{Comm}$), Community Hub-Bridge centrality ($\beta_{CHB}$), Participation Coefficient ($\beta_{PC}$), and Community-based Mediator centrality ($\beta_{CBM}$) behaves as the global community-aware centrality measures. Community-based centrality($\beta_{CBC}$), and K-shell with Community centrality ($\beta_{ks}$) behave like the local community-aware centrality measures.

GrQc is a network with a medium community structure strength ($\mu$ = 0.20). High variations of the correlation values are still visible for local community-aware centrality measures. For global community-aware centrality measures, globally, the correlation values are slightly higher. For the mixed community-aware centrality measures, two groups appear, similar to the ones before. However, they are now less pronounced.

 New Zealand Collaboration network has a weak community structure strength ($\mu$ = 0.564). In this case, most of the local community-aware centrality measures show a low correlation with classical centrality measures. It is notable that most of the global community-aware centrality measures are highly correlated with classical centrality measures, except for Betweenness and Leverage based on inter-community links ($\beta_b^G$, $\beta_{lev}^G$). Finally, one can still distinguish two groups for the mixed community-aware centrality measures. Nonetheless, the group that had low correlation with classical centrality measures in strong and weak community structured networks ($\beta_{Comm}$, $\beta_{CHB}$, $\beta_{PC}$, $\beta_{CBM}$) now only concerns Comm centrality ($\beta_{Comm}$), Community Hub-Bridge centrality ($\beta_{CHB}$). In other words, most of the mixed community-aware centrality measures now show a high correlation with classical centrality measures.

 % State that there are exceptions
Generally, networks in the same group have similar behavior. Nonetheless, there are few exceptions to the general trends reported for the typical networks. For example, the Mouse Visual Cortex (see Supplementary Fig. S10) network with a low mixing parameter value can be considered a network with a strong community structure ($\mu$ = 0.154). However, the local community-aware centrality measures show low correlation and not the global ones. The global and mixed community-aware centrality measures generally show a medium correlation. Caltech is another example (Supplementary Fig. S14) that departs from its class. Although it has a weak community structure ($\mu$ = 0.410), it shows a high correlation among almost all combinations between classical and community-aware centrality measures.
  
 % Comparison with LFR, must be later 
In comparison with artificial networks, generally, real-world networks comply with the behavior of synthetic networks. Indeed, in both cases, it appears that the community structure strength ($\mu$) is a major driver affecting the relationship between classical and community-aware centrality measures. For example, Ego Facebook, which is a strong community structured network ($\mu$ = 0.077), has a similar heatmap to the synthetic network generated with a strong community structure ($\mu$ = 0.05) illustrated in \autoref{SyntheticMainPaperBasicConfiguration}. %Too detailed: In addition, Community Hub-Bridge centrality ($\beta_{CHB}$) tends to show a low correlation with classical centrality measures in real-world networks. It is not the case with artificial networks. 
On the other extreme, New Zealand Collaboration (with its weak community structure $\mu$ = 0.564) has its global and mixed-community aware centrality measures that behave similarly to the synthetic network generated with a weak community structure ($\mu$ = 0.70). 
However, for local community-aware centrality measures, sometimes a lower correlation can be observed. The same observation is valid for other networks with a weak community structure (Kegg Metabolic, Internet Autonomous Systems, Interactome Vidal, and Human Protein).

This experiment shows that the relation between classical and community-aware centrality measures in real-world networks is greatly affected by the community structure strength. That is, for networks with strong community structure, local community-aware centrality measures exhibit a non-homogeneous behavior resulting in medium to high correlation values. Global community-aware centrality measures have a low correlation with classical centrality measures. Mixed community-aware centrality measures, split into two groups. The first group behaves as the global community-aware centrality measures. The second group shows a high correlation with classical centrality measures.  When the community structure strength weakens, the overall behavior gets inverted. In networks with a weak community structure, the global community-aware centrality measures show a high correlation with classical centrality measures. In contrast, the local community-aware centrality measures show a low correlation. Two groups still appear for the mixed community-aware centrality measures. However, the group exhibiting high correlation is the biggest.

\subsection*{Influence of the community detection algorithm } 
 This experiment aims to investigate the influence of the variation of the uncovered communities on the relationship between classical and community-aware centrality measures. To this aim, the community structure is uncovered using the Louvain algorithm\cite{blondel2008fast}. Comparisons are performed with the compression-based algorithm Infomap \cite{rosvall2008maps} outputs used in the previous experiments.
Louvain is a popular modularity-based community detection algorithm. Based on its output, the Kendall's Tau correlation between community-aware centrality measures and classical centrality measures is computed. In addition, the community structure strength ($\mu$) is also computed. The heatmaps and their respective community structure strength ($\mu$) for the 50 real-world networks are reported in the supplementary material. 

Globally, results show no fundamental difference with the overall behavior observed with Infomap. For example, let's consider the Ego Facebook network. It has a strong community structure based on Infomap ($\mu$ = 0.077) and Louvain ($\mu$ = 0.038). Both heatmaps show that the global community-aware centrality measures show a low correlation with classical centrality measures in this network. For local community-aware centrality measures, the correlation values fluctuate from low to high. Finally, mixed community-aware centrality measures are divided into two groups. The first group shows a low correlation with classical centrality measures, and the second exhibits a high correlation. The New Zealand Collaboration network is another typical example.  It has a weak community structure based on Infomap ($\mu$ = 0.564) and Louvain ($\mu$ = 0.412). Both heatmaps show how the global community-aware centrality measures are highly correlated with classical centrality measures while the local community-aware centrality measures show opposite behavior. Also, mixed community-aware centrality measures are divided into two groups. Note that correlation values may be higher or lower depending on the intensity of the community structure strength. For example, the correlation value between Degree centrality ($\alpha_d$) and Diffusion centrality based on intra-community links ($\beta_{dif}^L$) in New Zealand Collaboration using Infomap is 0.21. This value is lower than the one obtained with Louvain (0.15). It is not unexpected since Infomap extracts a weaker community structure.

Although the community structure strength's influence is comparable, one can notice three main differences when comparing the community detection algorithms. 
First, the values and, therefore, the range of the community structure strength differ. It ranges from $\mu$ = 0.077 to $\mu$ = 0.564 using Infomap while with Louvain it ranges from $\mu$ = 0.034 to $\mu$ = 0.712.

Second, as Louvain is fundamentally different from Infomap, it may uncover community structures that are nonidentical. Therefore, the estimated community structure strength can differ significantly. For example, the community structure revealed by Infomap in the 911AllWords network is a strong community structure ($\mu$ = 0.153) but a weak community structure ($\mu$ = 0.712) according to Louvain (refer back to Supplementary Fig. S10 and S24).
 
Third, a group of networks shows a low correlation on almost all possible combinations between classical and community-aware centrality measures using Infomap. It is not the case using Louvain, where a clear distinction can be made between local, global, and mixed community-aware centrality measures. These networks are London Transport, EuroRoad, and Internet Topology Cogentco. Hence, the community detection algorithm may sometimes mask the effect of community structure strength. Indeed, Infomap is able to identify more communities than Louvain \cite{lancichinetti2009community}. Accordingly, the intra-community and inter-community links using Infomap in some cases may be more heterogeneous for a given node than Louvain. Consequently, low correlation can be seen among classical and community-aware centrality measures.

%Too detailed + we removed fig27 of degree distributions: The Football network is an exception worth noting. Correlation between classical and community-aware centrality measures is low regardless of the community detection algorithm and the type of community-aware centrality measures (local, global, mixed). This network has a medium community structure strength ($\mu$ = 0.293). Going deeper, it appears that it is the only network with a left-skewed degree distribution, with skewness amounting to -1.343 (see Supplementary Fig. 27.). Hence, in this network, hubs are more frequent than low degree nodes. The low correlation between classical and community-aware centrality measures, in this case, can be attributed to this characteristic. The hubs may be considered influential by classical centrality measures. However, they may have low intra-community or inter-community links. As community-aware centrality measures use this information, a low correlation between classical centrality measures is observed.
 
To summarize, using different community detection algorithms can lead to a great variation of the uncovered community structure. This is particularly true with networks exhibiting a weak community structure. However, it does not significantly influence the behavior between classical and community-aware centrality measures as a function of the community structure strength.

\subsection*{Influence of the macroscopic and mesoscopic topological properties of the network}

This experiment investigates the relationship of the network topological properties with respect to the correlation between classical and community-aware centrality measures. Like in \cite{oldham2019consistency}, all pairs of correlation measures are quantified by their mean value. A high mean value indicates that, on average, community-aware centrality measures are highly correlated with classical centrality measures in a network. This value is then related to the network topological properties using linear regression. In fact, two mean values are calculated: the mean value for the local community-aware centrality measures and the mean value for the global community-aware centrality measures. Indeed, previous experiments have shown that they exhibit an opposite behavior. Correlation is high between classical and local community-aware centrality measures in networks with a strong community structure and low when networks have a weak community structure. One observes the contrary for global community-aware centrality measures. It is the reason why they are investigated separately. Although mixed community-aware centrality measurements can be classified into one of these two groups, they are discarded because their group can change across networks. Note that the mean values calculated in this experiment are based on the communities identified by Infomap.

The relationship between the means and the network topological properties is investigated using simple linear regression. Ordinary and weighted least squares estimators are used. In ordinary least squares, all the observations are weighted equally. In weighted least squares, an observation with high variance is weighted less than an observation with low variance during the fitting process. In the latter, weights are estimated using the approach of Wooldridge \cite{wooldridge2016introductory}. More details are provided in Supplementary Note 3.

In these models, the mean values are the dependent variables, and the topological properties of the networks are the independent variables. The relationship between the dependent and independent variables is considered statistically significant when the $p$-value is below a threshold value. There are seven macroscopic and nine mesoscopic topological features under test. The macroscopic features are Density, Transitivity, Assortativity, Average distance, Diameter, Efficiency, and Degree distribution exponent. Mixing parameter, Modularity, Internal distance, Internal density, Max-ODF, Average-ODF, Flake-ODF, Embeddedness, and Hub dominance are the mesoscopic properties. All these properties are representative structural characteristics of real-world networks and their communities \cite{barabasi2016network, lancichinetti2010characterizing, yang2015defining}. Their definitions are reported in Supplementary Note 2 and Supplementary Tables S1 and S2 contain their values for each network. Note that the exponent ($\gamma_{pred}$) of the power-law degree distribution is estimated using the maximum likelihood method \cite{clauset2009power}. The goodness of fit between the estimated distribution and the empirical distribution is computed using the Kolmogorov-Smirnov (KS) test. Among the fifty networks, Football is the only one that does not pass the test. Consequently, it is discarded in simple linear regression when studying the degree distribution exponent ($\gamma_{pred}$).

Simple linear regression allows investigating the relation between each topological property and the mean correlation values extracted for the local and global community-aware centrality measures. Results plotted in  \autoref{LinearRegUnivariate}  illustrate these relationships. Table S3 in supplementary reports significance, standard error, confidence interval, and coefficient of determination estimates using ordinary least squares.

% State significant macro+meso

Let us first consider the local community-aware centrality measures (\autoref{LinearRegUnivariate} (A)). The most significant macroscopic features are density, transitivity, and efficiency. Indeed, their $p$-value is low ($p \leq 0.01$). With a higher $p$-value ($p \leq 0.05$), average distance can also be considered to have a statistically significant linear relationship with the mean. It is not the case for assortativity, diameter, and the degree distribution exponent. Their high $p$-values allow concluding that there is no linear relationship with the mean. The mixing parameter and internal distance are the only mesoscopic characteristics exhibiting a significant relationship ($p \leq 0.05$). % Explain significant macro and meso
The mean correlation value between classical and local community-aware centrality measures increases with density, transitivity, and efficiency. Indeed, the denser the network, the more transitive, and the higher its efficiency (smaller shortest path distances between nodes). Increasing these parameters translates to a higher proportion of inter-community links. As a result, local community-aware centrality measures are drawing closer to classical centrality measures. On the contrary, the mean correlation value decreases with the average distance.  Indeed, as the average distance increases, nodes become less reachable, and local community-aware centrality measures behave more differently than classical centrality measures. There is a negative linear relationship between the mixing parameter and the mean correlation of classical and local community-aware centrality measures. It confirms the results of the experiments performed using artificial networks. The higher the mixing parameter, the lower the correlation between classical and community-aware centrality measures on average. One can also notice a negative linear relationship with internal distance. The higher the distance inside communities, the lower the correlation between local community-aware centrality measures and classical centrality measures.

\begin{figure}[htbp]
\begin{center}
\includegraphics[width=16cm,height=5.1cm]{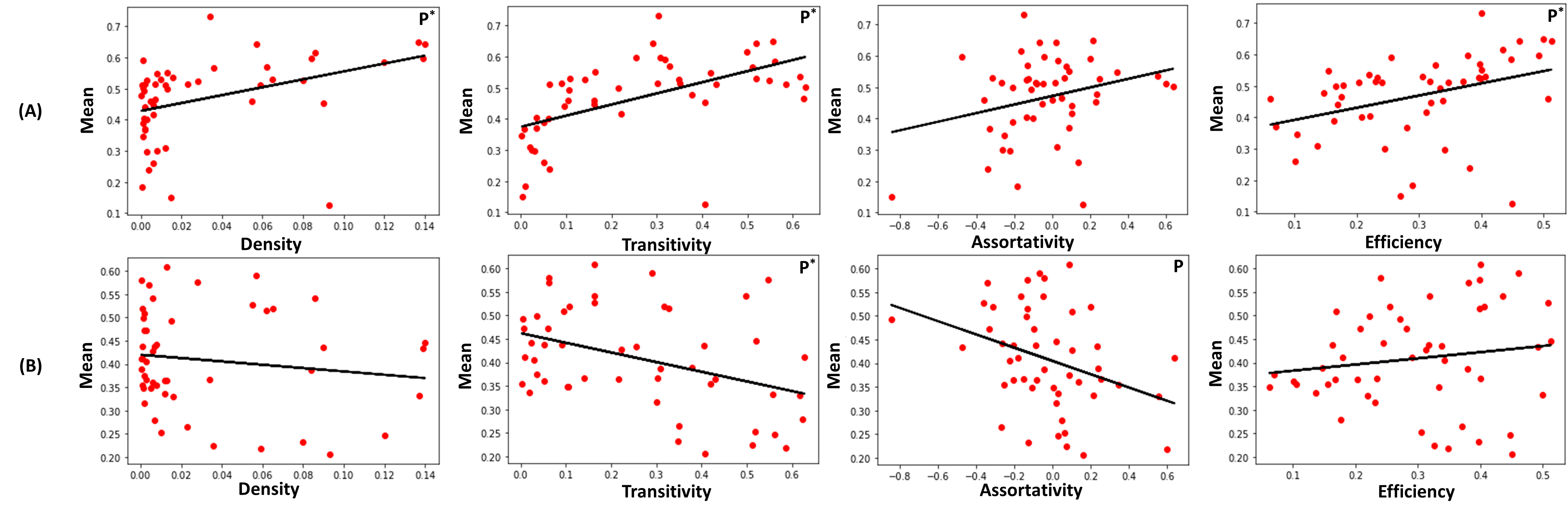}
\includegraphics[width=16cm,height=5.1cm]{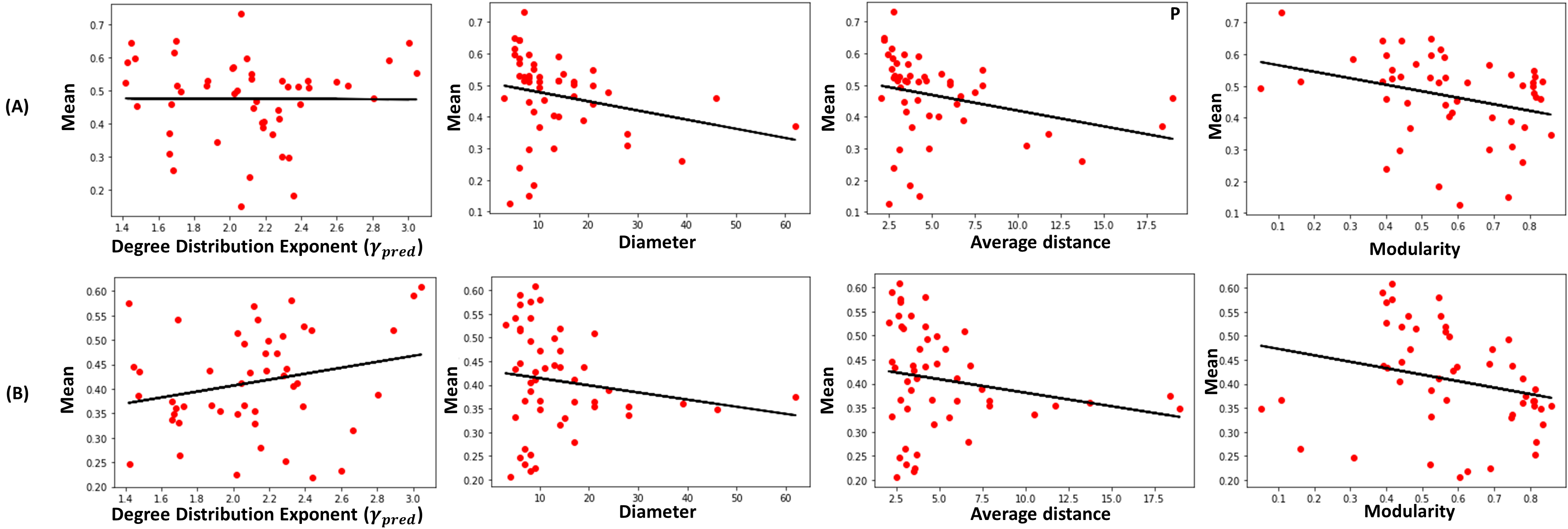}
\includegraphics[width=16cm,height=5.1cm]{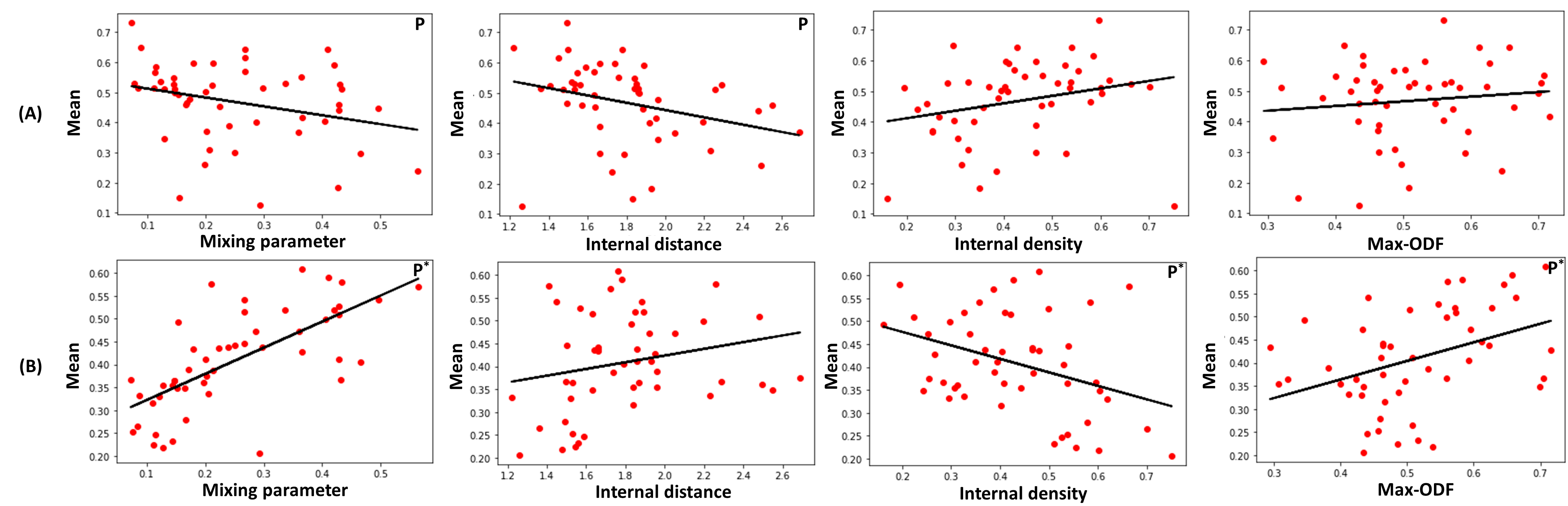}
\includegraphics[width=16cm,height=5.1cm]{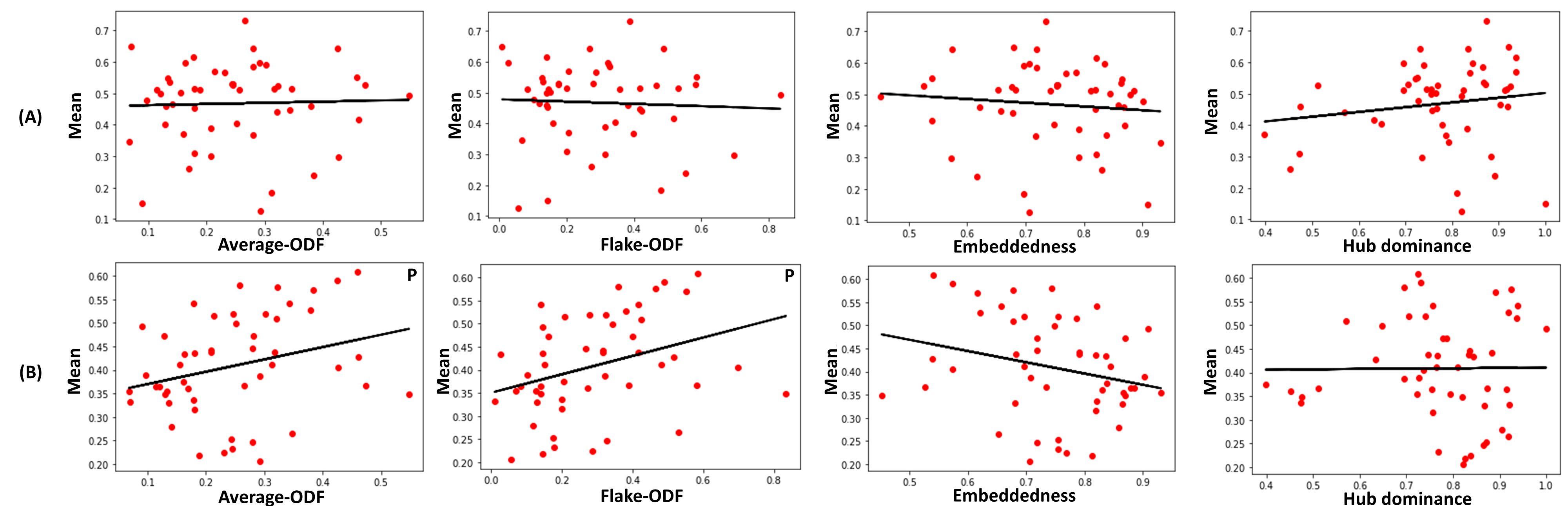}
   \centering
\caption{Relationship of the mean of the correlation between classical, local (A), and global (B) community-aware centrality measures with respect to the topological properties of real-world networks. The line is fitted by linear regression using ordinary least squares. "P" indicates $p\leq0.05$. "P" and * indicate $p\leq0.01$.
}
\label{LinearRegUnivariate}
\end{center}
\end{figure}

% State significant macro+meso

Let’s now turn to global community-aware centrality measures. The relationship between the mean correlation between classical and global community-aware centrality measures and macroscopic topological properties illustrated in (\autoref{LinearRegUnivariate} (B))  is weaker. Transitivity is the only property with a small $p$-value ($p \leq 0.01$). Although weaker, the linear relationship between assortativity and the mean correlation is statistically significant ($p \leq 0.05$). However, compared with the local community-aware centrality case, more mesoscopic characteristics are involved in a significant relationship with the mean correlation value. The mixing parameter, internal density, and Max-ODF exhibit a significant relationship ($p \leq 0.01$), followed by Flake-ODF and Average-ODF ($p \leq 0.05$). % Explain significant macro and meso
As transitivity increases, the mean correlation value between classical and global community-aware centrality measures decreases. It may be because the more transitive a network, the higher the proportion of intra-community links to inter-community links. Hence, the correlation between global community-aware and classical centrality measures decreases as the former relies inter-community links and the latter does not distinguish these two types of connections. Assortativity exhibits similar behavior. An increase in assortativity indicates that nodes tend to be more attached to peers. A hub in a community may be likely linked to a hub in another community, while small degree nodes are also likely to connect to low degree nodes in other communities. As a result, global community-aware centrality measures are more capable than classical centrality measures to distinguish nodes in this situation. An increase in the mixing parameter increases the correlation between classical and community-aware centrality measures. The effect of the mixing parameter confirms the results of the previous experiment. On a strong community structure (low mixing parameter), the mean of the correlation between classical and global community-aware centrality measures is low. As the community structure becomes weaker (higher mixing parameter), the correlation's mean starts to increase. 
There is a negative linear relationship between internal density and the mean correlation of classical and global community-aware centrality measures. The denser the communities, the lower the correlation between classical and global community-aware centrality measures. Indeed, in this case, communities act as sub-networks of their own. Hence, inter-community links provide the unique distinctiveness of nodes inside dense communities as classical centrality measures account for indistinguishable links. There is a positive relationship between Max-ODF and the mean correlation. Indeed, it is related to the nodes acting as bridges in their communities. The more a node has inter-community links, the more connected to other communities, the less distinctive the inter-community links are from the other links in a network. Average-ODF and Flake-ODF behave similarly to Max-ODF.

% Commenting on WLS
The estimates using weighted least squares generally confirm the previous results of ordinary least squares concerning the relationship between classical and local community-aware centrality measures with some additional significant features (see Supplementary Table S4 for details). The relationship between density, transitivity, efficiency, average distance, mixing parameter, internal distance, and the mean correlation is still statistically significant. In addition, the diameter, modularity, hub dominance become also significant ($p \leq 0.05$). The diameter has a negative linear relationship with the mean correlation. The higher the network's diameter, the further the distance between the nodes, and the further classical and local community-aware centrality measures disagree.  Modularity has a negative linear relationship as well. In a highly modular network, community-aware centrality measures may diverge from classical ones. The linear relationship between Hub dominance and the mean correlation is positive. An increase in the hub size means an increase in its intra-community links with respect to all of its links in its community. As a result, the hubs become less heterogeneous, and the correlation of local community-aware centrality measures with classical centrality measures gets higher.
 %(?????Hub dominance measures the fraction of intra-community links of each hub in each community?????). 

Using weighted least squares instead of ordinary least squares in the estimation process of global community-aware centrality measures does not change the results fundamentally. The linear relationship of average distance ($p \leq 0.05$) and modularity ($p \leq 0.01$) with the mean correlation becomes statistically significant, and it is no more the case for Average-ODF. The association is negative for the two features. An increase in modularity widens the distinction between classical and local as well as global community-aware centrality measures. It results in a lower correlation in both cases. Similar behavior is observed for the average distance.

To sum up, using the overlapping results between ordinary and weighted least squares, density, transitivity, efficiency, average distance, mixing parameter, and internal distance play a significant role in shaping up the correlation between classical and local community-aware centrality measures. While the most significant topological properties influencing the relationship between classical and global community-aware centrality measures are transitivity, assortativity, mixing parameter, internal density, Max-ODF, and Flake-ODF.

\section*{Conclusion}
 
This work investigates the interplay between classical, community-aware centrality measures, and network topology. It relies on an extensive empirical analysis of artificial networks with controlled topological properties and real-world networks originating from various domains.
Synthetic network analysis shows that the community structure strength significantly impacts the correlation between classical and community-aware centrality measures. One can distinguish two groups of community-aware centrality measures that behave inversely.
The more well-separated the communities, the more the global community-aware centrality measures decorrelate from classical centrality measures. Inversely, the more local community-aware centrality measures correlate with classical centrality. Furthermore, the degree distribution exponent and the community size distribution exponent have almost no influence on this relationship.
Results with real-world networks are consistent with synthetic networks regarding the impact of the community structure strength on the relation between community-aware and classical centrality measures. Using a different detection algorithm can lead to a different community structure but it doesn't change the relationship between the community structure strength and the relationship between classical and community-aware centrality measures. Linear regression analysis shows that in any case, the relationship between classical and community-aware centrality measures is highly sensitive to transitivity. Density, average distance, efficiency, and internal distance are also important parameters influencing the relation between classical and local community-aware centrality measures. Assortativity, internal density, Max-ODF, and Flake-ODF turn out to be the most influential properties affecting the relationship between classical and global community-aware centrality measures.
This work allows us to understand better how network topology, classical centrality measures, and community-aware centrality measures interact. It shows that community-aware centrality measures are more distinct in situations where the community structure is strong. In future work, we plan to assess community-aware centrality measures' effectiveness in various information diffusion scenarios.

%\bibliographystyle{unsrt}  


\end{document}